\documentclass[preprint,12pt]{elsarticle}

\usepackage{amssymb}
\usepackage{amsmath}
\usepackage{epsfig}
\usepackage{graphicx}
\usepackage{float}
\usepackage{url}
\usepackage[labelsep=period]{caption}
\usepackage{subcaption}
\usepackage{multirow}
\usepackage{amsfonts,amsthm}
\usepackage{mathtools}
\usepackage{commath}
\usepackage[sc,osf]{mathpazo}
\usepackage[linesnumbered,ruled]{algorithm2e}
\usepackage[super]{nth}
\usepackage{verbatim} % comments
\usepackage{pgfplots, pgfplotstable}
\usepackage{pgffor}
\usepackage{parskip}

\pgfplotsset{compat=1.18}

\newcommand{\DT}{{I}}                  % Domaine temporel
\newcommand{\vx}{\vect{x}} % x vector

\newcommand{\PGDm}{m} % Mode number notation.

\newcommand{\p}{\hspace{0.5mm}} % producto entre matrices y vectores.

\newcommand{\vect}[1]{\underline{{#1}}} 

\newcommand{\SDvect}[1]{%
            \underline{\boldsymbol{
                {#1}
            }}
}

\newcommand{\SDtens}[1]{%
            \underline{\underline{\boldsymbol{
                {#1}
            }}}
}
 
\newcommand{\normi}[1]{\left\|{#1}\right\|}

\journal{Signal Processing}

\begin{document}

\begin{frontmatter}



\title{Single Atom Convolutional Matching Pursuit: Theoretical Framework and Application to Lamb Waves based Structural Health Monitoring}


%\author{} %% Author name

\author[label1]{Sebastian Rodriguez \corref{mycorrespondingauthor}}
\cortext[mycorrespondingauthor]{Corresponding author}
\ead{sebastian.rodriguez_iturra@ensam.eu}
\author[label1]{Marc Rébillat}
\author[label1]{Shweta Paunikar}
\author[label1]{Pierre Margerit}
\author[label1]{Eric Monteiro}
\author[label1,label2,label3]{Francisco Chinesta}
\author[label1]{Nazih Mechbal}

\affiliation[label1]{organization={PIMM, Arts et Métiers ParisTech, CNRS, CNAM},%Department and Organization
            addressline={151 Boulevard de l'Hôpital}, 
            city={Paris},
            postcode={75013}, 
            state={},
            country={France}}

\affiliation[label2]{organization={ESI Group Chair @ Arts et Métiers Institute of Technology},%Department and Organization
            addressline={151 Boulevard de l'Hôpital}, 
            city={Paris},
            postcode={75013}, 
            state={},
            country={France}}

\affiliation[label3]{organization={CNRS@CREATE LTD},%Department and Organization
            addressline={1 Create Way, \#08-01 CREATE Tower}, 
            city={Singapore},
            postcode={138602}, 
            state={},
            country={Singapore}}

%% Abstract
\begin{abstract}
Structural Health Monitoring (SHM) aims to monitor in real time the health state of engineering structures. For thin structures, Lamb Waves (LW) are very efficient for SHM purposes. A bonded piezoelectric transducer (PZT) emits LW in the structure in the form of a short tone burst. This initial wave packet (IWP) propagates in the structure and interacts with its boundaries and discontinuities (and in particular with eventual damages) generating additional wave packets. The main issues with LW based SHM are that at least two LW modes are simultaneously excited by a single PZT and that those modes are dispersive, i.e., that various frequencies propagate at various speeds. Measured signal by the other PZTs thus corresponds to propagated versions of the IWP that have interacted with structural discontinuities, and isolating additional echoes caused by damages is a key point for SHM. Matching Pursuit Method (MPM), which consists of approximating a signal as a sum of different delayed and scaled atoms taken from an a priori known learning dictionary, seems very appealing in such a context. MPM is, however, limited to non-dispersive signals and relies on a priori known learning dictionary containing candidates' atoms. An improved version of MPM called the Single Atom Convolutional Matching Pursuit method (SACMPM), which addresses the dispersion phenomena by decomposing a measured signal as delayed and dispersed atoms and limits the learning dictionary to only one atom, is proposed here. After describing the theoretical framework allowing the numerical setting up of SACMPM, its performances are illustrated when dealing with numerical and experimental LW-based SHM signals. It is then shown that the provided SACMPM decomposition is extremely efficient when coupled with machine learning algorithms for LW-based damage localization, thus demonstrating its practical interest regarding SHM. Although the signal approximation method proposed in this paper finds an original application in the context of SHM, this method remains completely general and can be easily applied to any signal processing problem.
\end{abstract}

\begin{graphicalabstract}
\end{graphicalabstract}

\begin{highlights}
\item Signal approximation by a Single Atom Matching Pursuit Method (SAMPM)
\item Signal approximation by a Single Atom Convolutional Matching Pursuit Method (SACMPM)
\item Damage identification using Neural Networks and the features extracted from SAMPM and SACMPM
\end{highlights}

%% Keywords
\begin{keyword}
Lamb Waves \sep Convolutional Matching Pursuit \sep Matching Pursuit \sep Structural Health Monitoring \sep Single Atom Dictionary


\end{keyword}

\end{frontmatter}



\section{Introduction}
\label{sec:Introduction}

\subsection{Structural Health Monitoring of thin Structures using Lamb waves}

In industrial applications, one of the major engineering challenges is to monitor structural damage in near real time, automatically, in order to prevent catastrophic failure, and this process is known as structural health monitoring (SHM) \citep{su_identification_2009, sohn_review_2002}. A classic SHM procedure generally consists of five stages \citep{sohn_review_2002, yuan_structural_2016, deraemaeker_new_2011}: detection, localization, classification, quantification, and prognostic. The term \emph{"damage"} is used here to define changes in the material properties and/or geometry of these structures, including boundary conditions, which have a negative effect on its current or future performance \citep{worden_fundamental_2007}. 

Among the various existing SHM techniques dedicated to thin composite or metallic structures, strategies based on ultrasonic Lamb Waves (LW) emitted and received by piezoelectric elements (PZT) are particularly effective \citep{su_guided_2006, su_identification_2009, mitra_guided_2016, qing_piezoelectric_2019}. LW are bending and compression waves (also called A0 and S0 modes in their lower frequency range) that stress the entire thickness of the thin structure being monitored. These waves have the particularity of being able to propagate over relatively large distances and can therefore cover a large control surface with few PZTs in a short time. However, LW possesses two main drawbacks: at any given frequency, at least two modes simultaneously coexist (namely A0 and S0), and these modes are dispersive, meaning that LW velocities depends on the frequency, which makes the interpretation of the collected signals tricky in practice.

The basic idea underlying LW-based SHM is then to excite one PZT bonded on the structure to monitor with a tone burst signal centered around a given frequency. This initial wave packet (IWP) then propagates through the structure to be inspected and interacts with its boundaries, its structural discontinuities, and eventual damages. Each of these discontinuities produces an additional wave packet propagating in the host structure. Consequently, the resulting signals measured by the other PZT correspond to the IWP after propagation within the host structure and multiple interactions caused by structural discontinuities. In particular, LW based SHM algorithms seek to detect echoes caused by the presence of damage in such signals in order to infer damage presence, location, type, and severity.

\subsection{LW based SHM signal decomposition strategies}

In such a context, being able to decompose measured signals as many wave packets that can be physically interpreted and potentially linked to structural damages is of great interest. Several signal processing methods have already been proposed to address this issue.

Matching Pursuit Method (MPM) \cite{mallat1993matching} proposes to approximate a given signal $s(t)$ as follows:
\begin{equation*}
s(t) \approx 
 \sum_{i=1}^{\PGDm} \alpha_{i} \Psi_{i}(t)
\end{equation*}
The features extracted by MPM correspond to the atoms $\Psi_{i}(t)$ and their amplitudes $\alpha_{i}$. Typically the atoms $\Psi_{i}(t)$ are selected on an over-complete learning dictionary $D$ that needs to be provided a priori \citep{xu2009lamb,raghavan2007guided,chakraborty2009damage,lu2008numerical,mu2021ultrasound}, where functions $\Psi_{i}(t)$ are selected in a greedy process in the sparsest way possible. However, MPM is limited only to non-dispersive signals, preventing its application in SHM for thin structures where LW undergoes dispersion during propagation. In this case, in addition to delays and attenuation, wave packets also endure dispersion caused by the fact that all the frequencies do not propagate at the same speed within the structures under study. Furthermore, the learning dictionary $D$ needs to be a priori defined, which limits the practical use of MPM as not all delayed, attenuated, and dispersed atoms cannot be precomputed in advance for both A0 and S0 modes propagating in a continuous thin structure.

 Despite these drawbacks, MPM has, however, already been applied in an LW based SHM context in combination with other methods for damage monitoring. Recently, Li et al. \citep{li_damage_2024} developed a method using orthogonal matching pursuits and model updating to locate damages in hinge joints of a fully functional hollow slab bridge using data collected from a single location. To decompose and reconstruct the various wave packets in a signal, Gao et al. \citep{gao_defect_2023} used orthogonal matching pursuit algorithm based on dictionary matrix, while Kim and Yuan \citep{kim_adaptive_2020} used it for imaging damage in an aluminium plate, and Li et al. \citep{li_orthogonal_2019} used it to reconstruct normalized electromechanical admittance data to develop an SHM system for a concrete tunnel. Mu et al. \citep{mu_ultrasound_2021} also used orthogonal MP in conjunction with dispersion removal for identifying LW packets in shells. MPM is employed by Hong et al. \citep{hong_damage_2019} to analyze guided wave signals obtained from FE simulations and experiments in the covered region of the metallic messenger cable in an electrified railway catenary and successfully detect damage larger than 2.5 mm in the cable. An optimized dictionary based MPM was employed to study and size axial defects in in-service or corroded pipelines by Tse and Wang \citep{tse_characterization_2013}. It is noteworthy that all of these methods are based on predefined over-complete dictionaries for employing MPM.
 
Other non-dictionary based approaches like Variational Mode Decomposition (VMD), Empirical Mode Decomposition (EMD), Proper Orthogonal Decomposition, etc., have also been used by researchers to decompose wave signals in different domains. Cuomo et al. \citep{cuomo2023damage} applied Hilbert Huang transform based EMP to decompose a multi-component signal and detect impact damage in aluminium and CFRP plates. A modified two parameters based VMD method was used by Jaiang et al. \citep{jiang2023quantitative} to identify wave modes in multi-mode ultrasonic wave signals and further detect and quantify internal holes of various sizes at various depths in rail specimens. Another version of VMD, which extracts the mode functions successively using four pre-defined decomposition criteria, is used by Zeng et al. \citep{zeng2022two} to identify faults in long electrical transmission cables. However, these non-dictionary based approaches are based on a mathematical basis and not on any physical basis, thus limiting the interpretability of the provided results in practice.

\subsection{Proposed approach}

An improved version of Matching Pursuit is thus proposed in the present work. It addresses the decomposition of LW based SHM signals by decomposing a measured signal as delayed and dispersed impulse response of a single atom. This decomposition is called here the \emph{Single Atom Convolutional Matching Pursuit Method} (SACMPM). First, a theoretical framework is presented for the computation of MPM where a purely mathematical construction of the decomposition is achieved without the need of an over-represented learning dictionary which is called here the  \emph{Single Atom Matching Pursuit Method} (SAMPM). Here, only a single atom $\Psi(t)$ is used, which is considered to be the external excitation imposed on the structure and corresponds to the initial wave packet. A theoretical framework allowing to numerically obtain the optimal amplitudes and time delay of a SAMPM decomposition is presented following a \emph{Greedy} process building the decomposition on-the-fly until convergence. Then, its extension to the SACMPM, taking into account dispersion effects through a convolution operation, is introduced on the basis of the previous theoretical framework. Both methods are afterward applied to experimental LW based SHM signals for comparison purposes and to highlight the benefits offered by the SACMPM. Finally, damage localization is achieved through the use of machine Learning algorithms fed by features extracted from SAMPM and SACMPM thus demonstrating its practical interest for SHM purposes.

It should be noted that the signal approximation methods proposed in this paper (SAMPM and SACMPM) remain completely general and can be easily applied to any signal processing problem.

The present paper is thus structured as follows: Sec.~\ref{sec:classic_MP} introduces the \emph{Single Atom  Matching Pursuit Method} (SAMPM) and the proposed associated theoretical framework. Sec.\ref{sec:MP_conv} then extends it to \emph{Single Atom Convolutional Matching Pursuit} (SACMPM). Following Sec.~\ref{sec:Efficiency} and Sec.~\ref{sec:Damage_detect} provide numerical analysis on the performance of the proposed techniques when they approximate real signals and when they are used to predict damage location, respectively. Finally,  Sec.~\ref{sec:Concl_Pers} provides conclusions and perspectives.

\section{Single Atom Matching Pursuit Method theoretical framework}\label{sec:classic_MP}

The Single Atom Matching Pursuit Method (SAMPM) is introduced here along with the theoretical framework, allowing to numerically build the optimal decomposition without the need for a priori known learning dictionary. 

\subsection{Selecting a single initial atom}
\label{sec:atom_det}

\begin{figure}[!ht]
\centering
    \includegraphics[width=\textwidth]{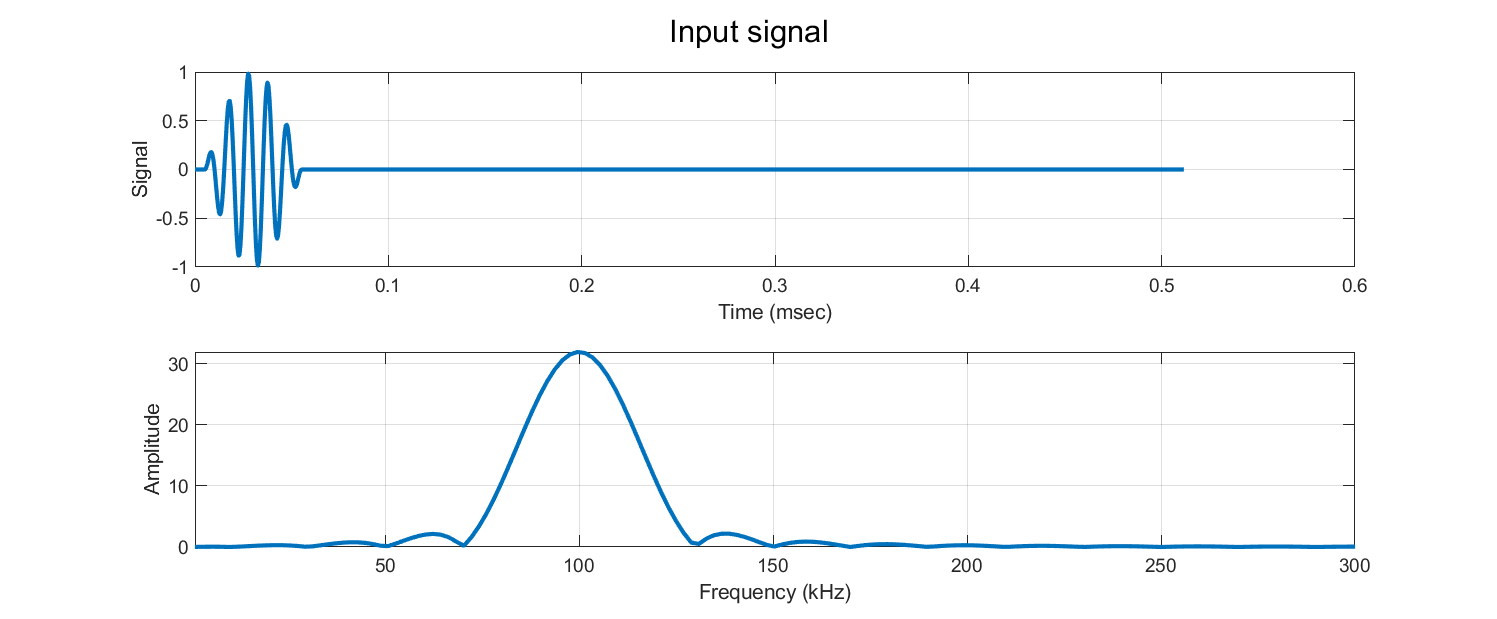}
    \caption{Typical input signal used for LW SHM purposes. Here the central frequency $f_0$ is $100$~kHz and the burst is composed of $5$ cycles with a half-sinusoidal window.}
    \label{fig:input}
\end{figure}

For LW-based SHM applications, the excitation signal is known and corresponds to the initial wave packet (IWP) being sent on the structure. This signal is usually a tone burst centered around a given central frequency, as shown in Fig.\ref{fig:input}. As this IWP is the origin of all the upcoming echoes appearing in the host structure and being later measured, it can naturally be used as the only atom contained in the learning dictionary. The main assumption made in that case is that this single IWP is at the core of all the other generated wave packets and thus can be used as the single atom that is necessary in the learning dictionary. As the signal being sent to PZT is systematically recorded in LW-based SHM applications, this single atom is readily available in practice, and thus, there is no need for any a priori known learning dictionary in the present case.

\subsection{Frequency domain greedy approach}

The main idea behind SAMPM then consists in approximating a given measured signal $s(t)$ after its propagation within the structure to be inspected by the following decomposition:
\begin{equation}\label{eq:decomposition}
    s(t) \approx s_{\PGDm}(t) =  \sum_{i=1}^{\PGDm} \alpha_{i} \Psi(t-\tau_{i})
\end{equation}
where the principal unknowns to be determined by the MPM algorithm correspond to the amplitude of each term $\alpha_{i}$ and its temporal delay $\tau_{i}$ and the only considered atom $\Psi(t)$ corresponds to the one that has been previously selected. It is proposed here that each term of the decomposition is determined one after the other in an incremental way, in a so-called \emph{greedy} process. 

Let's suppose the decomposition is known until $\PGDm-1$ terms, such that one can write:
\begin{equation}
    s_{\PGDm}(t) = s_{\PGDm-1}(t) + \alpha_{\PGDm} \Psi(t-\tau_{\PGDm})
\end{equation}
where the remaining unknown correspond to $\alpha_{\PGDm}$ and $\tau_{\PGDm}$.  These unknowns are determined in a way such that they minimize an error between the approximation and the reference signal. This error can be defined in time as follows:
\begin{equation}\label{eq:min_prob_time}
\lbrace \alpha_{\PGDm}, \tau_{\PGDm} \rbrace = \underset{\lbrace \alpha_{\PGDm}, \tau_{\PGDm} \rbrace}{\text{arg min}}
    \normi{ \alpha_{\PGDm} \Psi(t-\tau_{\PGDm}) - s^{\PGDm-1}_{res}(t) }_{\DT_{t}}^{2}
\end{equation}
where $s^{\PGDm-1}_{res}(t)$ corresponds to the residual signal, and is therefore given as follows:
\begin{equation}
s^{\PGDm-1}_{res}(t) = s(t) - s_{\PGDm-1}(t)  
\end{equation}
and the used norm is defined as follows: 
\begin{equation*}
\normi{\cdot}^2_{\DT_t} = \int\limits_{I_{t}} (\cdot)^2 dt
\end{equation*}
with the temporal domain $\DT_{t} = [0,T]$. 

The minimization problem can also be defined in the frequency domain using the Parseval's theorem:
\begin{equation}\label{eq:min_prob_freq}
\lbrace \alpha_{\PGDm}, \tau_{\PGDm} \rbrace = \underset{\lbrace \alpha_{\PGDm}, \tau_{\PGDm} \rbrace}{\text{arg min}}
    \normi{ \alpha_{\PGDm} \hat{\Psi}(\omega) e^{-j \omega \tau_{\PGDm}} - \hat{s}^{\PGDm-1}_{res}(\omega) }_{\DT_{\omega}}^{2}
\end{equation}
with $\hat{\bullet} = \mathcal{F}(\bullet)$ (Fast Fourier Transform) and its corresponding norm defined in the frequency domain:
\begin{equation} 
\normi{ \cdot }_{\DT_{\omega}}^{2} =  \int_{\DT_{\omega}} ( \cdot )^{*} ( \cdot ) d\omega
\end{equation}
where $( \cdot )^{*}$ corresponds to the complex conjugate of $( \cdot )$.

\subsection{Optimal determination of amplitudes $\alpha_{\PGDm}$ and phases $\tau_{\PGDm}$}\label{sec:amplitude_phases}

A simple numerical algorithm is presented here, allowing obtaining the optimal amplitudes and phases for the term $\PGDm$ of the decomposition in the sense of Eq.~\eqref{eq:min_prob_freq}. For this, first let's notice that by minimizing the functional of Eq.~\eqref{eq:min_prob_freq} with respect to the amplitude $\alpha_{\PGDm}$ by employing variational calculus \cite{troutman2012variational}, one obtains:

$\forall \delta\alpha \in \mathbb{R}$,
\begin{equation*}
\int\limits_{\DT_{\omega}}  \delta\alpha_{\PGDm} \left(\hat{\Psi}(\omega) e^{-j \omega \tau_{\PGDm}} \right)^{*} \left( \alpha_{\PGDm} \hat{\Psi}(\omega) e^{-j \omega \tau_{\PGDm}} - \hat{s}^{\PGDm-1}_{res}(\omega)  \right) d\omega = 0
\end{equation*}
which directly implies:
\begin{equation}\label{eq:amplitude_gen}
\alpha_{\PGDm}(\tau_{\PGDm}) =  \frac{ \text{Re} \left( \int\limits_{\DT_{\omega}} \left( \hat{\Psi}(\omega) e^{-j \omega \tau_{\PGDm}} \right)^{*} \hat{s}^{\PGDm-1}_{res}(\omega) d\omega \right) }{ \int\limits_{\DT_{\omega}} \left( \hat{\Psi}(\omega) e^{-j \omega \tau_{\PGDm}} \right)^{*} \left( \hat{\Psi}(\omega) e^{-j \omega \tau_{\PGDm}} \right) d\omega }  
\end{equation}
where $\text{Re}(\bullet)$ represent here the real part of $\bullet$. Equation \eqref{eq:amplitude_gen} gives the optimal amplitude by minimizing \eqref{eq:min_prob_freq} with respect to any value of $\tau_{\PGDm}$. The real part function is used in Eq.~\eqref{eq:amplitude_gen} since  $\alpha_{\PGDm}(\tau_{\PGDm})$ is a real number.

\begin{comment}
As $\alpha_{\PGDm}(\tau_{\PGDm})$ is a real number and as numerical procedure may still produce a small but negligible imaginary part, it is safer in practice to take the real part of Eq.~\eqref{eq:amplitude_gen}.
\end{comment}

Now that we know for a given delay $\tau_{\PGDm}$ the optimal amplitude $\alpha_{\PGDm}(\tau_{\PGDm})$ to retain, let's minimize the error norm with respect to the time phase $\tau_{\PGDm}$. To do so, let's first notice that by developing \eqref{eq:min_prob_freq}, one obtains:
\begin{equation}
\begin{split}\label{eq:norm_freq_dev0}
&\normi{ \alpha_{\PGDm} \hat{\Psi}(\omega) e^{-j \omega \tau_{\PGDm}} - \hat{s}^{\PGDm-1}_{res}(\omega) }_{\DT_{\omega}}^{2} =  \\ & 
\normi{\alpha_{\PGDm} \hat{\Psi}(\omega) e^{-j \omega \tau_{\PGDm}}}_{\DT_{\omega}}^{2} - g(\alpha_{\PGDm},\tau_{\PGDm})  + \normi{\hat{s}^{\PGDm-1}_{res}(\omega)}_{\DT_{\omega}}^{2} 
\end{split}
\end{equation}
with:
\begin{equation*}
g(\alpha_{\PGDm},\tau_{\PGDm}) = 2 \alpha_{\PGDm} \text{Re} \left( \int\limits_{\DT_{\omega}} \left( \hat{\Psi}(\omega) e^{-j \omega \tau_{\PGDm}} \right)^{*} \hat{s}^{\PGDm-1}_{res}(\omega) d\omega \right) 
\end{equation*}
If one injects the expression of the amplitude $\alpha_{\PGDm}(\tau_{\PGDm})$ \eqref{eq:amplitude_gen} that minimizes the error norm, into expression \eqref{eq:norm_freq_dev0} one obtains after simplification:
\begin{equation}\label{eq:norm_freq_dev}
\normi{ \alpha_{\PGDm} \hat{\Psi}(\omega) e^{-j \omega \tau_{\PGDm}} - \hat{s}^{\PGDm-1}_{res}(\omega) }_{\DT_{\omega}}^{2} =  
\normi{\hat{s}^{\PGDm-1}_{res}(\omega)}_{\DT_{\omega}}^{2} - G( \tau_{\PGDm} )  
\end{equation}
where:
\begin{equation}
G(\tau_{\PGDm}) = \frac{ \left[ \text{Re} \left( \int\limits_{\DT_{\omega}} \left( \hat{\Psi}(\omega) e^{-j \omega \tau_{\PGDm}} \right)^{*} \hat{s}^{\PGDm-1}_{res}(\omega) d\omega \right) \right]^2 }{ \int\limits_{\DT_{\omega}} \left( \hat{\Psi}(\omega) e^{-j \omega \tau_{\PGDm}} \right)^{*} \left( \hat{\Psi}(\omega) e^{-j \omega \tau_{\PGDm}} \right) d\omega }  
\end{equation}
From \eqref{eq:norm_freq_dev}, one can clearly recognize that the value of $\tau_{\PGDm}$ that minimizes the error simply corresponds to the one that maximizes $G(\tau_{\PGDm})$. The optimal value of $\tau_{\PGDm}$ can be easily computed as $G(\tau_{\PGDm})$ is a one-dimensional function. From the optimal value of the time phase $\tau_{\PGDm}$, the amplitude $\alpha_{\PGDm}$ is finally determined via \eqref{eq:amplitude_gen}.

\subsection{Convergence and the stop criterion}

The procedure of computing each of the terms of equation \eqref{eq:decomposition} is performed until a given approximation error is reached. This error is defined as follows:
\begin{equation}
\xi_{\PGDm} = 100 \times \frac{\normi{\sum_{i=1}^{\PGDm} \alpha_{i} \Psi(t-\tau_{i}) - s(t)}_{\DT_t}}{\normi{s(t)}_{\DT_t}}
\end{equation}

\section{Single Atom Convolutional Matching Pursuit}\label{sec:MP_conv}

In order to better approximate the dispersion phenomena observed on signals under the SAMPM rationale proposed in the previous Section, we propose here to include a convolution operation, allowing taking into account the dispersion effect on the selected initial atom during its propagation over the structure to monitor.

\subsection{Problem statement}

Here, we propose the following decomposition:
\begin{equation}\label{eq:conv_MP}
    s(t) \approx s_{\PGDm}(t) = \sum_{i=1}^{\PGDm} \left[ \alpha_{i}(t) * \Psi(t-\tau_{i}) \right](t)
\end{equation}
where $*$ corresponds to the convolution operator, which involves the following:
\begin{equation}
(f*g)(t) = \int\limits_{-\infty}^{\infty} f(s)g(t-s) ds
\end{equation}
Thus, from the above definition, if we denote $\Psi_{\tau_{i}}(t) = \Psi(t-\tau_i)$ (the input signal $\Psi(t)$ delayed by the quantity $\tau_{i}$), the operation in \eqref{eq:conv_MP} simply consists of:
\begin{equation*}
\left[ \alpha_{i}(t) * \Psi(t-\tau_{i}) \right](t) = \int\limits_{-\infty}^{\infty} \alpha_i(s)\Psi_{\tau_{i}}(t-s) ds     
\end{equation*}
On this proposed decomposition, one considers no longer a scalar amplitude $\alpha_\PGDm$ multiplying the delayed atom but rather a temporal signal $\alpha_\PGDm(t)$ being convolved with the delayed atom. Furthermore, this temporal signal $\alpha_\PGDm(t)$ can be interpreted as an impulse response relating the $\PGDm^\text{th}$ wave packet to the delayed initial atom due to the fact that a convolution operation is used between $\alpha_\PGDm(t)$ and the atom $\Psi(t-\tau)$. Consequently, the family of $\alpha_\PGDm(t)$ will called the wave packets impulse responses. In this work, the proposed decomposition is called Single Atom Convolutional Matching Pursuit method (SACMPM).

\subsection{Wave packets impulse responses approximation}

Here the wave packets impulse responses are approximated by using Chebyshev polynomials of the second kind. Therefore one has:
\begin{equation}\label{eq:temp_conv}
    \alpha_{\PGDm}(t) = \sum_{i=1}^{N} N_{i}(t) \beta_{i} = \SDvect{N}(t)^{T} \SDvect{\beta}
\end{equation}
where, $\beta_{i}$ and $N_{i}(t)$ correspond to the weight (of each shape function) and the shape function, respectively. This choice of approximation of the time function is made due to the simplicity of the implementation and improved conditioning of the operators needed to be inverted.

\textbf{Remark}: Here, because the used atom has a local support in time (interval size where the function is defined), the function $\alpha(t)$ has consequently this same interval of definition.

\subsection{Greedy construction of the decomposition}
Following the same strategy as the one exposed for SAMPM, here, the wave packet impulse responses $\alpha_{i}(t)$ and the time-delays of the atom $\tau_{i}$ are computed in a greedy way. In this sense, let's suppose the decomposition known until $\PGDm-1$ terms, such as we can write:
\begin{equation}
    s_{\PGDm}(t) = s_{\PGDm-1}(t) + \left[ \alpha_{\PGDm}(t) * \Psi(t-\tau_{\PGDm}) \right](t)
\end{equation}
therefore, the main unknowns are computed by minimizing the following norm:
\begin{equation}\label{eq:min_prob_time_conv}
\lbrace \alpha_{\PGDm}, \tau_{\PGDm} \rbrace = \underset{\lbrace \alpha_{\PGDm}, \tau_{\PGDm} \rbrace}{\text{arg min}}
    \normi{ \left[\alpha_{\PGDm}(t)*\Psi(t-\tau_{\PGDm}) \right](t) - s^{\PGDm-1}_{res}(t) }_{\DT_{t}}^{2}
\end{equation}
with $s^{\PGDm-1}_{res}(t) = s(t) - s_{\PGDm-1}(t)$.

\subsection{Determination of temporal time delays $\tau_{\PGDm}$}

The temporal delays are determined exactly in the same way as exposed in Sec.~\ref{sec:amplitude_phases} by assuming a constant amplitude $\alpha_{\PGDm}$. The main reason motivating this choice is based on the fact that obtaining an optimal solution of wave packet impulse responses and time delay is really complicated based on iterative resolution methods applied to Eq.~\eqref{eq:min_prob_time_conv}, due to ill-conditioning of the operators needed to be inverted, in addition, the iterative procedure is completely eliminated since $\tau$ is computed directly.

\subsection{Determination of the wave packet impulse responses $\alpha_{\PGDm}(t)$ given $\tau_{\PGDm}$}

Given $\tau_{\PGDm}$, the temporal function $\alpha_{\PGDm}(t)$ is computed such as it minimizes \eqref{eq:min_prob_time_conv}. By minimizing with respect to this function \cite{troutman2012variational}, it results:

$\forall \delta\alpha_{\PGDm}(t)$,
\begin{equation}\label{eq:min_dev_conv}
\int\limits_{\DT_t} \left[ \delta\alpha_{\PGDm}(t)*\Psi(t-\tau_{\PGDm}) \right](t) \left( \left[ \alpha_{\PGDm}(t)*\Psi(t-\tau_{\PGDm})\right](t) - s^{\PGDm-1}_{res}(t) \right) dt = 0    
\end{equation}
By introducing \eqref{eq:temp_conv} into \eqref{eq:min_dev_conv}, one obtains the following matrix equation:
\begin{equation*}
\SDtens{M} \p \SDvect{\beta} = \SDvect{F}    
\end{equation*}
with the matrix $\SDtens{M}$ and vector $\SDvect{F}$ given by:
\begin{equation}\label{eq:matrix_SACMPM}
\SDtens{M} = \int\limits_{\DT_t} \SDvect{B}(t) \p \SDvect{B}(t)^{T} dt \quad \text{and} \quad \SDvect{F} = \int\limits_{\DT_t} \SDvect{B}(t) \p s^{\PGDm-1}_{res}(t) dt
\end{equation}
%
%
\begin{comment}
\begin{equation}
\SDvect{F} = \int\limits_{\DT_t} \SDvect{B}(t) \p s^{\PGDm-1}_{res}(t) dt
\end{equation}
\end{comment}
%
%
where the shape functions are given by:
\begin{equation}
\SDvect{B}(t) = \left[\SDvect{N}(t) * \Psi(t-\tau_{\PGDm})\right](t) 
\end{equation}

\subsection{Convergence and the stop criterion}

New terms are added to the decomposition until a given approximation error is reached. This error is defined as follows:
\begin{equation}
\xi_{\PGDm} = 100 \times \frac{\normi{\sum_{i=1}^{\PGDm} \left[\alpha_{i}(t) * \Psi(t-\tau_{i})\right](t) - s(t)}_{\DT_t}}{\normi{s(t)}_{\DT_t}}
\end{equation}

\section{Efficiency of SAMPM and SACMPM applied to numerical and experimental LW-based SHM signals}\label{sec:Efficiency}

In this Section, the previously presented decomposition algorithms (namely SAMPM and SACMPM see Sec.~\ref{sec:classic_MP} and Sec.~\ref{sec:MP_conv}) will be applied to approximate signals which are representative of LW based SHM applications. Firstly, numerical signals corresponding to LW propagation in an infinite isotropic plate will be considered in Sec.~\ref{sec:TH}. Then SAMPM and SACMPM will be applied to experimental signals collected on the fan cowl part of an A380 nacelle in Sec.~\ref{sec:XP}.

\subsection{Numerical example}
\label{sec:TH}

\subsubsection{Simulated signals to be approximated}
As a first step to demonstrate the decomposition abilities of the SAMPM and SACMPM algorithms a very simple numerical case is considered. This case consists of the propagation of $A_0$ and $S_0$ modes in an infinite thin plate, and details regarding the simulation are provided in~\ref{sec:numerical_example}. 
\begin{figure}[!ht]
\centering
    \includegraphics[width=0.75\textwidth]{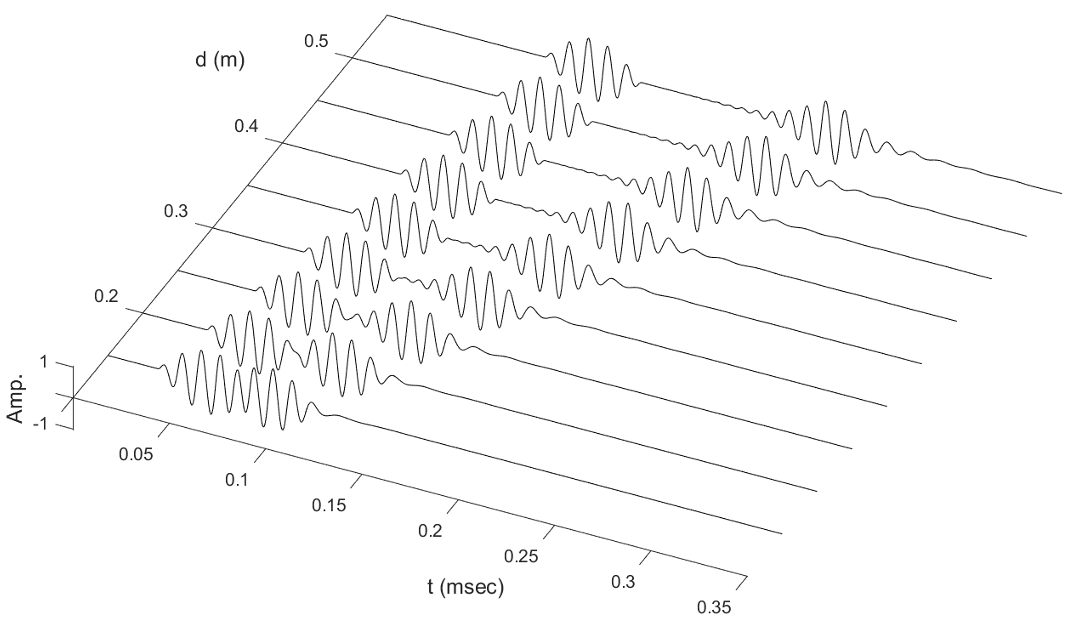}
    \caption{Simulated signals for various propagating distances.}
    \label{fig:propTH}
\end{figure}
The numerical signal consist of $1024$ samples, with a sampling frequency of $2$~MHz, and correspond to an input having a central frequency $f_0=100$~kHz and with $5$~cycles bursts. Signals have been computed for distances $d$ ranging from $15$~cm to $55$~cm by step of $5$~cm. Such distances are quite common in practice for the LW based SHM of aeronautic composite structures, as will be shown later. The resulting signals corresponding to these computations are shown in Fig.~\ref{fig:propTH}. As can be seen from this figure, the signals are made up of 2 wave packets: the first one, the fastest, corresponds to $S_0$, and the second one, the slowest, to $A_0$. As expected from the dispersion curves analysis shown in~\ref{sec:numerical_example}, the $S_0$ wave packet does not suffer from dispersion, whereas the $A_0$ one is distorted during propagation because of dispersion.

\subsubsection{Comparison of SAMPM and SACMPM methods performances}

The SAMPM and SACMPM algorithms described in Sec.~\ref{sec:classic_MP} and Sec.~\ref{sec:MP_conv} have then been applied to the simulated signals described previously with an initial atom corresponding to the input signal existing the structure. For the SACMPM method, the discretization parameter $N$  has been set to $40$. A maximum of $50$ terms of the decomposition has been allowed, and convergence was supposed to be reached when the error was lower than $10$~\%. One should keep in mind that as two wave packets are physically propagating here, the minimum number of terms that could be found by either SAMPM or SACMPM is $2$.
\begin{figure}[!ht]
\centering
    \includegraphics[width=\textwidth]{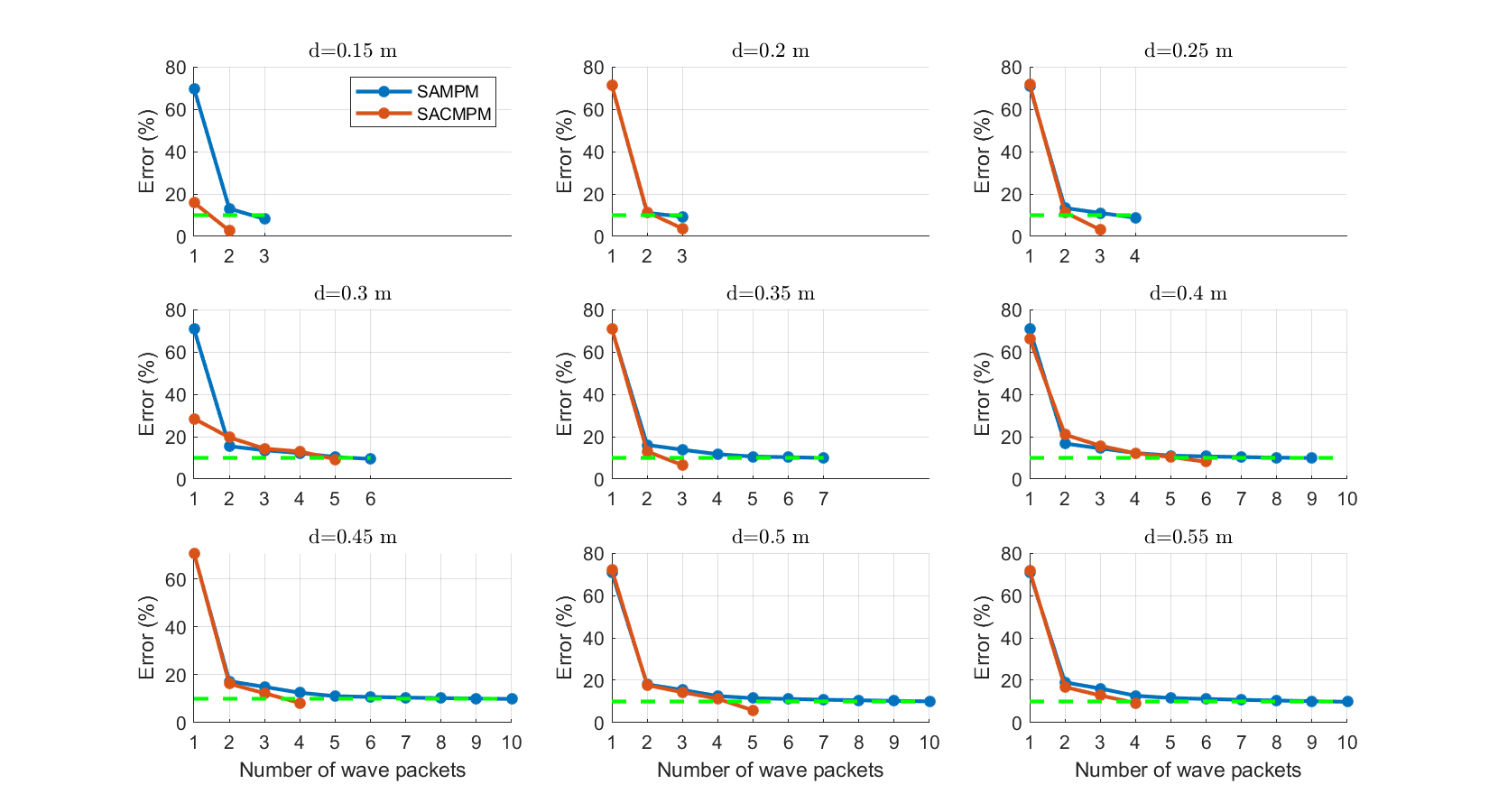}
    \caption{Comparison of the performances of the SAMPM and SACMPM algorithms for the simulated signals corresponding to different propagation distances.}
    \label{fig:perfcomp}
\end{figure}
A comparison of the performances of the SAMPM and SACMPM algorithms for the simulated signals corresponding to different propagation distances is shown in Fig.~\ref{fig:perfcomp}. From this figure, it can be observed that using the input signal as the initial atom both algorithms converge to an error lower than $10$~\% with $3$ and $4$ terms for propagation distance corresponding to $15$~cm, $20$~cm, and $25$~cm. According to Fig.~\ref{fig:propTH}, for those propagation distances, the two wave packets to be identified are either mixed ($15$~cm) or very close to each other ($20$~cm and $25$~cm). Results provided by SAMPM and SACMPM are thus acceptable. For larger distances ($d\geq 25$~cm), SACMPM converges with fewer terms than SAMPM. In those cases, the wave packets to be identified are well separated according to Fig.~\ref{fig:propTH}. The fact that SAMPM needs more terms to converge is related to the fact that this algorithm is not able to learn, dispersion whereas SACMPM can.
\begin{figure}[!ht]
\centering
    \includegraphics[width=\textwidth]{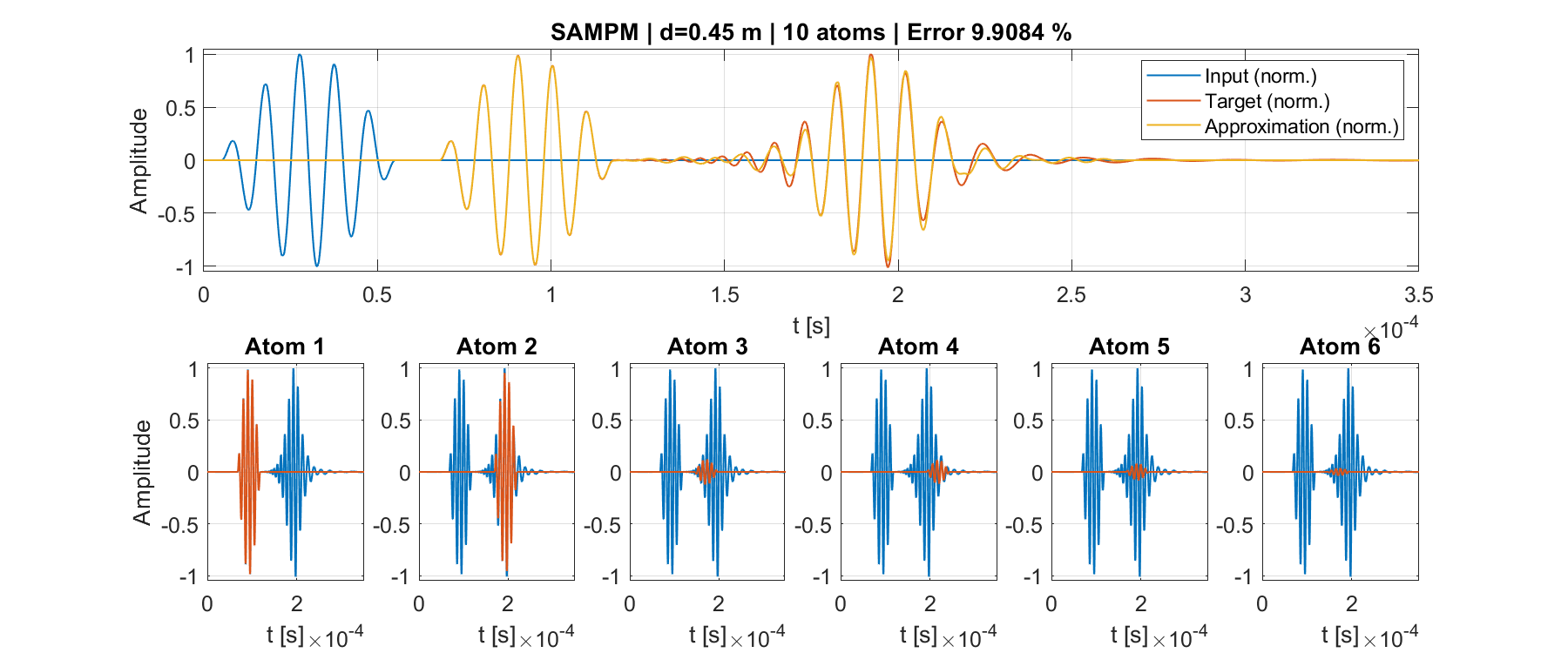}
    \caption{\emph{[Top]} Input signal, target signal, and approximated signal using the SAMPM algorithm with 10 terms. \emph{[Bottom]} First 6 terms obtained using the SAMPM for the signal corresponding to a propagation distance of $45$~cm. Convergence with a $10$~\% error was not reached with three terms, but ten terms are required.}
    \label{fig:perfcompSAMPM}
\end{figure}
The approximated signal as well as the first 6 terms obtained using the SAMPM  for the signal corresponding to a propagation distance of $45$~cm are shown in Fig.~\ref{fig:perfcompSAMPM}. From this figure, it can be seen that for an error lower than $10$~\%, the approximation provided by SAMPM is extremely satisfying. It can also be seen that the first term of the decomposition corresponds exactly to the propagated $S_0$ mode. This was to be expected as the SAMPM algorithm just propagates the input signal without dispersion and as the $S_0$ mode does not endure dispersion. The second term being identified by the SAMPM algorithm corresponds, however, roughly to the second propagating term. This was also expected as this term is a propagated version of the input signal but with some dispersion effects. Consequently, it resembles the input signals but with some differences that can clearly be seen here. The terms $3$, $4$, and $5$ then seek to correct the small mismatches between the translated input signal and the actually observed second wave packet. Finally, the term number $6$ corrects some very small amplitude mismatches obtained in the first wave packet.
\begin{figure}[!ht]
\centering
    \includegraphics[width=\textwidth]{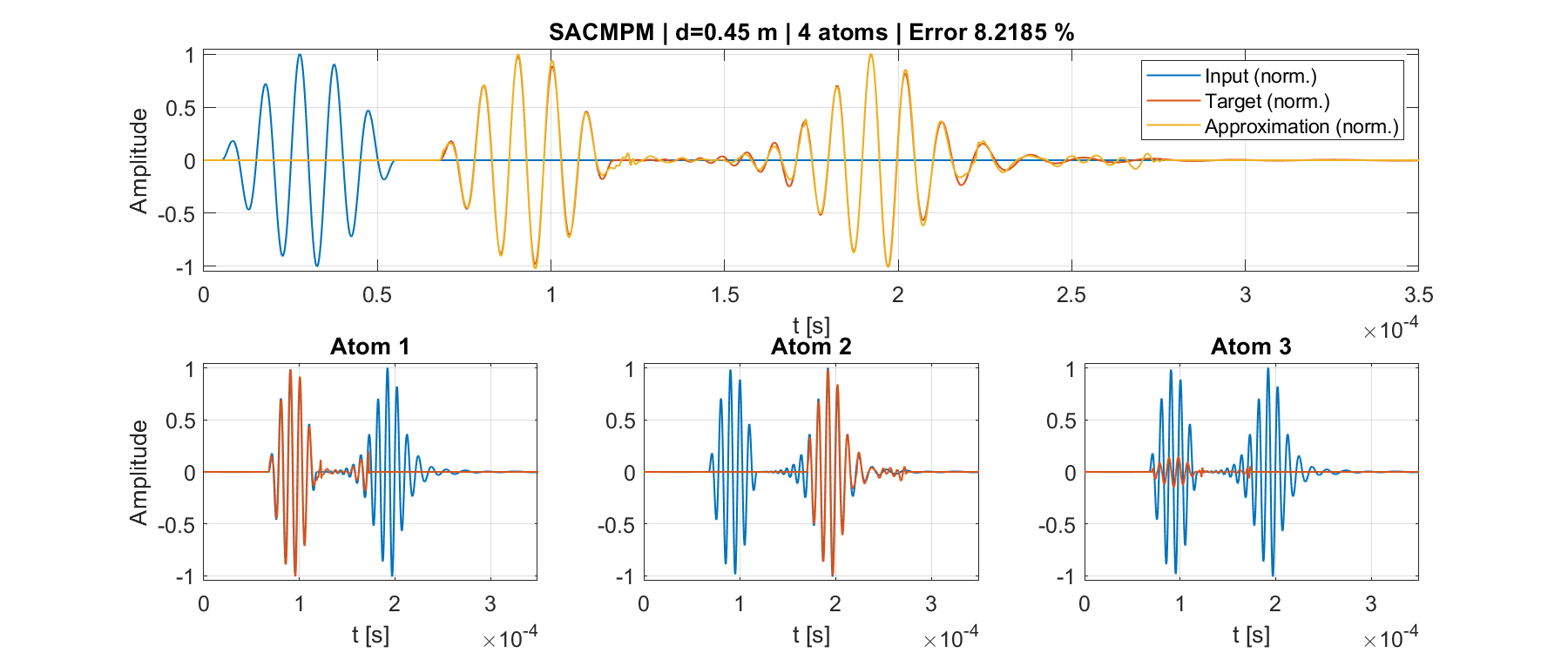}
    \caption{\emph{[Top]} Input signal, target signal, and approximated signal using the SACMPM algorithm with 4 terms. \emph{[Bottom]} First 3 terms obtains using the SACMPM for the signal corresponding to a propagation distance of $45$~cm. Convergence with a $10$~\% error was reached with four terms.}
    \label{fig:perfcompSACMPM}
\end{figure}
The approximated signal, as well as the first $3$ terms obtained using the SACMPM for the signal corresponding to a propagation distance of $45$~cm are then shown in Fig.~\ref{fig:perfcompSACMPM}. Again, the approximation provided by SACMPM is extremely satisfying. Furthermore, the first term almost perfectly corresponds to the first wave packet to approximate, as expected. The second term also almost perfectly corresponds to the second wave packets. Now that convolution has been included in the SACMPM algorithm, it is possible to better represent the dispersive behavior of the signal, as illustrated here.

%\begin{comment}

\subsubsection{Overview}

In summary, it is demonstrated here in this simple simulated example, which is representative of the targeted LW based SHM application, but includes only two wave packets that both the SAMPM and the SACMPM algorithms are able to efficiently approximate the target signals. Furthermore, as the SACMPM algorithm allows for compensation of the dispersion effects, it is more efficient than the SAMPM algorithm in doing so.

%\end{comment}

\subsection{Experimental example}
\label{sec:XP}

\subsubsection{Considered experimental data: fan cowl of the A380 Fan Cowl Structure}

\begin{figure}[!ht]
\centering
\includegraphics[height=4.5cm]{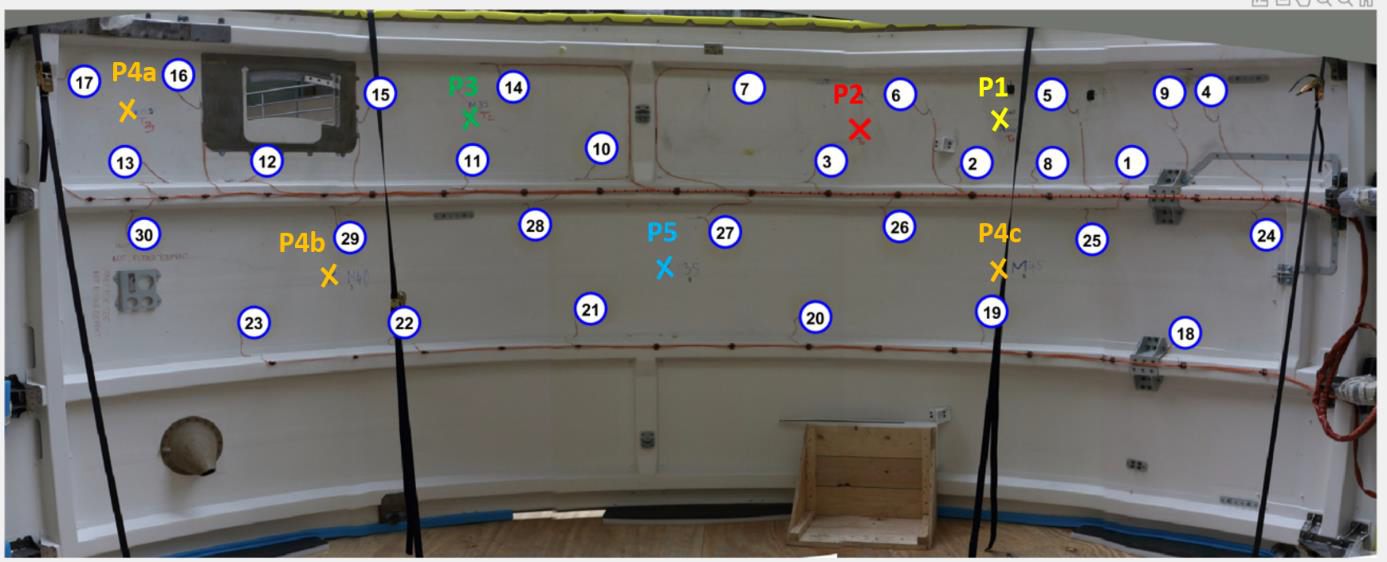}
    \caption{Overview of the geometrical configuration of the Fan Cowl Structure (FC) experimental datasets}
    \label{fig:FC}
\end{figure}

\begin{figure}[!ht]
\centering
    \includegraphics[width=0.75\textwidth]{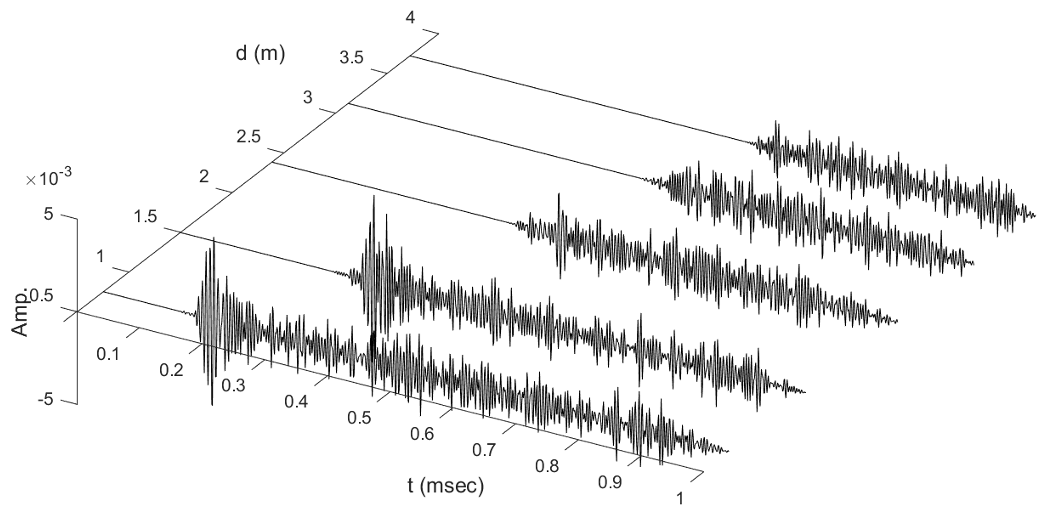}
    \caption{Experimental signals for various propagating distances.}
    \label{fig:XPsigs}
\end{figure}

The aeronautics structure under study experimentally consists here in the fan cowl part of a nacelle of an Airbus A380 as shown in Fig.~\ref{fig:FC}. The experimental details related to the structure under study are given in \ref{sec:XP_details}.
The considered experimental signals are shown in Fig~\ref{fig:XPsigs}. From this figure, it can be observed that the experimental signals are not as simple as the simulated ones previously presented. Indeed, in addition to the two initially propagating modes $A_0$ and $S_0$, reflections coming from structural boundaries and inhomogeneities such as the stiffeners (see Fig.~\ref{fig:FC}) generate additional wave packets that are also caught by the receiving piezoelectric elements. One can also notice that due to damping present in composite materials, the amplitude of the first peak decreased with increasing propagating distance. This structure has already been used by the authors for other studies, and more details can be found in related works~\cite{rebillat2018peaks,rebillat2020damage,Guo2022b}.

\subsubsection{Comparison of SAMPM and SACMPM methods performances}

\begin{figure}[!ht]
\centering
    \includegraphics[width=\textwidth]{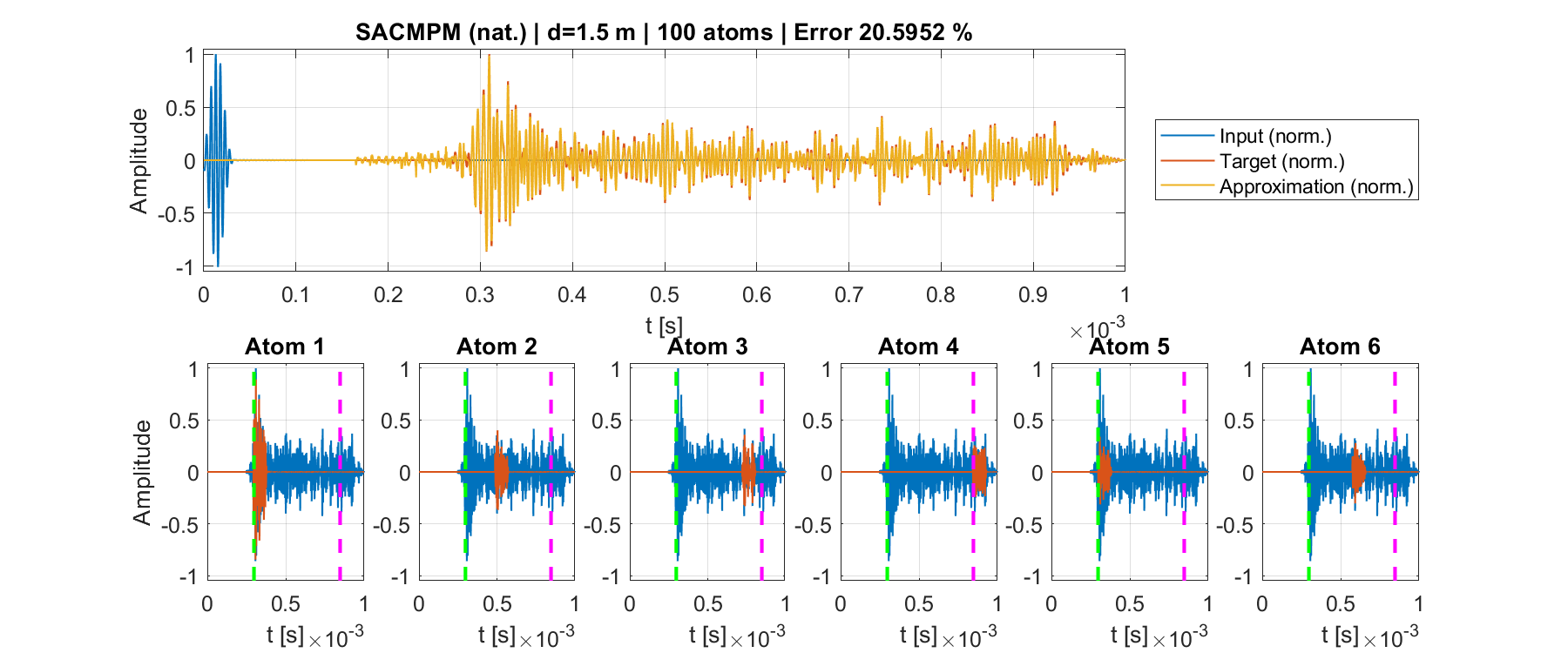}
       \caption{\emph{[Top]} Input signal, target signal, and approximated signal using the SAMPM algorithm with 100 terms. \emph{[Bottom]} First 6 terms obtained using the SACMPM for the signal corresponding to a propagation distance of $1.5$~m. The green and magenta vertical lines denote the expected arrival times of the waves packets corresponding to the $S_0$ and $A_0$ modes.}
    \label{fig:perfcompSACMPM_XP}
\end{figure}

The SAMPM and SACMPM algorithms described in Sec.~\ref{sec:classic_MP} and Sec.~\ref{sec:MP_conv} have then been applied to the experimental signals described previously. As previously, for the SACMPM method, the discretization parameter $N$  has been set to $40$. A maximum of $100$ terms has been allowed to approximate the experimental signals.
\begin{figure}[!ht]
\centering
    \includegraphics[width=\textwidth]{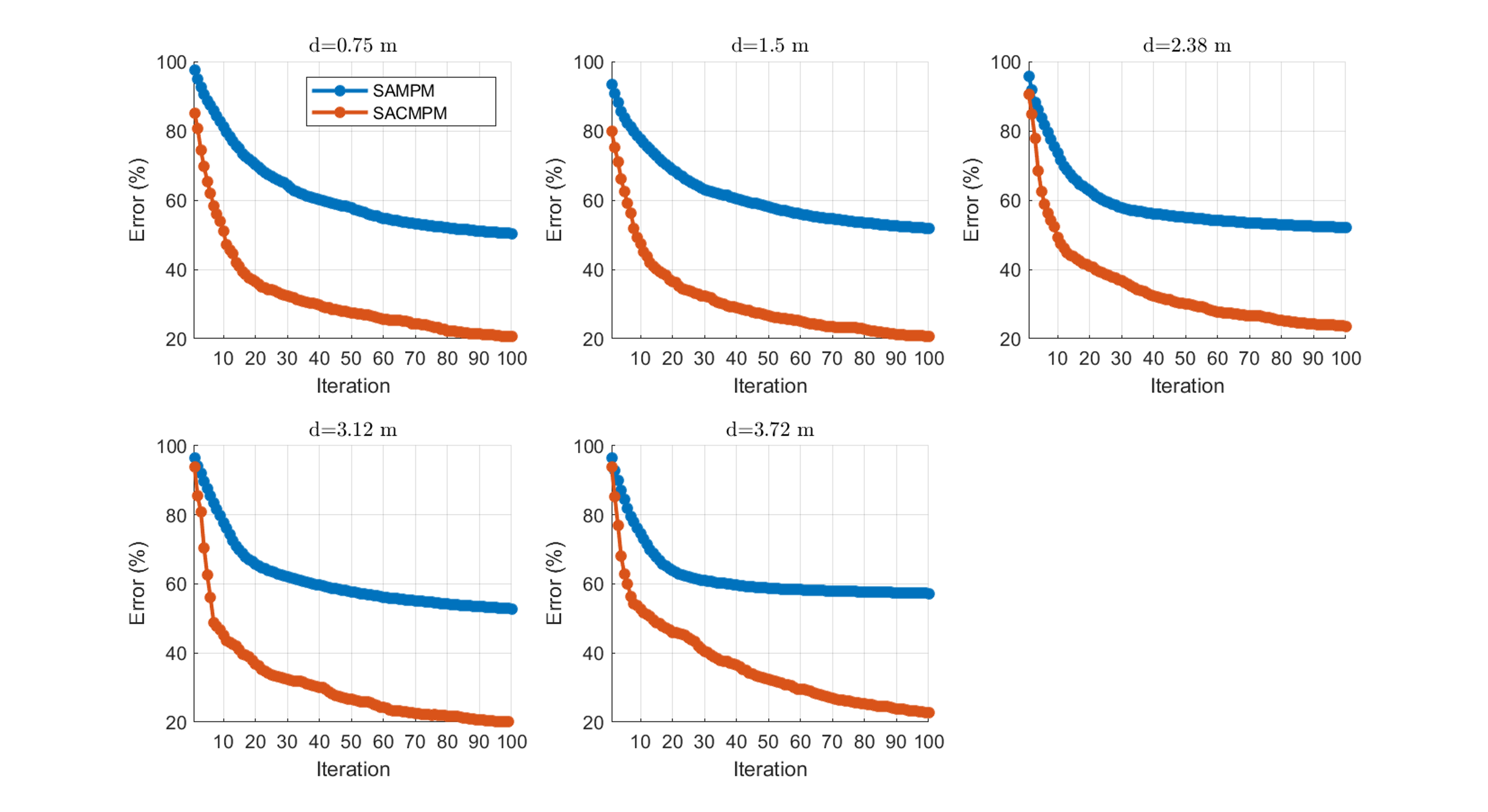}
    \caption{Comparison of the performances of the SAMPM and SACMPM algorithms for the experimental signals corresponding to different propagation distances.}
    \label{fig:perfcompXP}
\end{figure}
A comparison of the performances of the SAMPM and SACMPM algorithms for the experimental signals corresponding to different propagation distances are shown in Fig.~\ref{fig:perfcompXP}. From this figure, it can be observed that given $100$ terms, SAMPM converges to an error of roughly $50$~\%, whereas SACMPM converges to an error of $\simeq20$~\% for all distances. In that case, convergence is faster for SACMPM than for SAMPM for all the tested distances. This can be again interpreted by the fact that SACMPM is able to take into account dispersion during its estimation process, whereas SAMPM cannot. Convergence curves associated with SAMPM are, however, smoother than the ones of SACMPM. This can be attributed to the fact that the way SAMPM is being solved always ensures that the optimal amplitude and the best delay for a given atom are found. For SACMPM, some bumps can be observed in the convergence curves; this is mainly due to the fact that the time function $\alpha(t)$ (which is convolved with the atom) is built on a basis, which may penalize the approximation a little if the signal to be processed has a rich and complex frequency content. However, its global approximation for a fixed number of terms in the decomposition is always guaranteed to be better than SAMPM by construction.
\begin{figure}[!ht]
\centering
    \includegraphics[width=\textwidth]{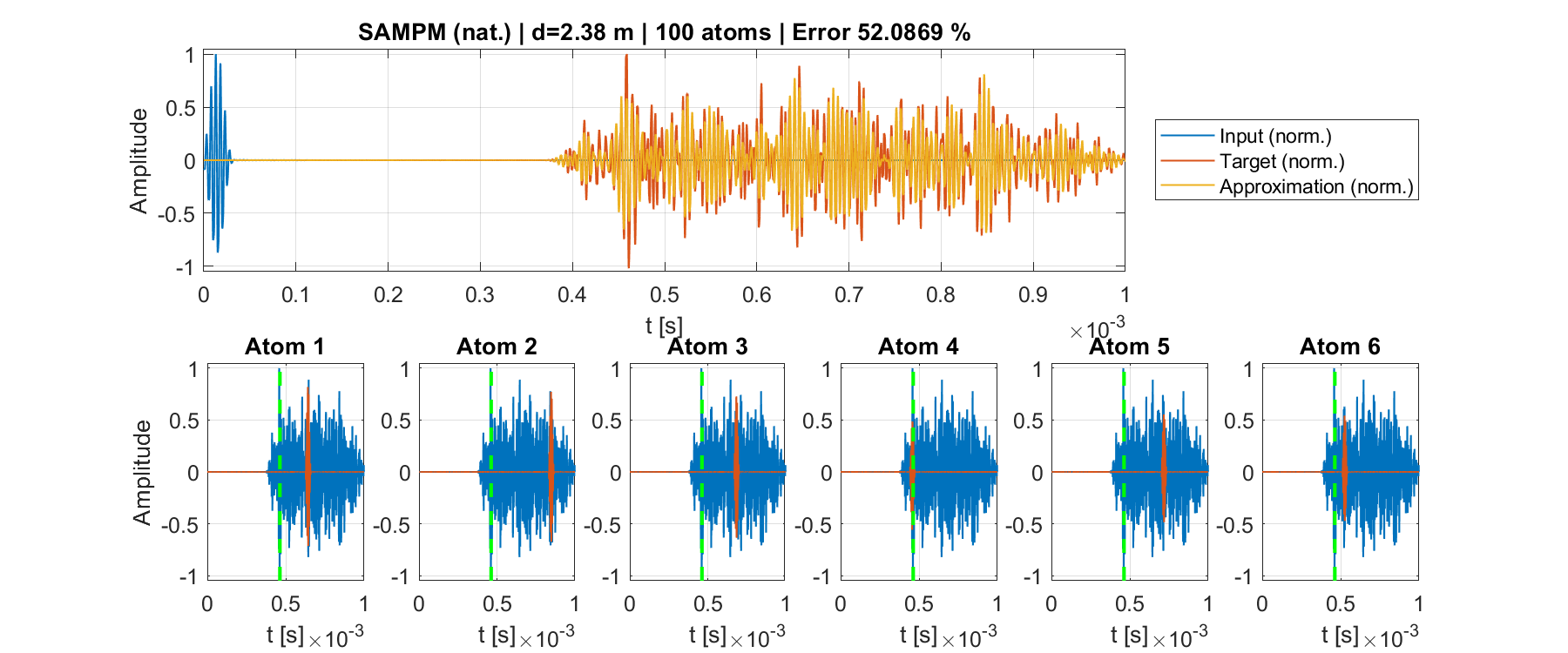}
    \caption{\emph{[Top]} Input signal, target signal, and approximated signal using the SAMPM algorithm with 100 terms. \emph{[Bottom]} First 6 terms obtains using the SAMPM for the signal corresponding to a propagation distance of $1.5$~m. The green and magenta vertical lines denote the expected arrival times of the waves packets corresponding to the $S_0$ and $A_0$ modes.}
    \label{fig:perfcompSAMPM_XP}
\end{figure}
The approximated signal, as well as the first 6 terms obtained using the SAMPM for the signal corresponding to a propagation distance of $1.5$~m are then shown in Fig.~\ref{fig:perfcompSACMPM_XP}. The approximation provided by SAMPM is satisfactory given the complexity of signals at hand, even if the error at the end is of $\simeq50$~\%. Interpretation of the terms being found is here, however, trickier given the complexity of the structure at hand. 
The approximated signal, as well as the first 6 terms obtained using the SACMPM for the signal corresponding to a propagation distance of $1.5$~m are then shown in Fig.~\ref{fig:perfcompSACMPM_XP}. Again, the approximation provided by SACMPM is very satisfying given the complexity of signals at hand, even if the error at the end is of $\simeq20$~\%. Interpretation of the terms being found is, here again, quite tricky, given the complexity of the structure at hand. The first and fifth terms seem to be related to the wave packet corresponding to the $S_0$ mode. Interestingly, the third and fourth wave packets seem to be associated with the $A_0$ mode wave packet. This suggests that on these experimental signals, SACMPM allows us to better catch the physics behind the analyzed signals than SAMPM.

%\begin{comment}

\subsubsection{Overview}

In summary, the analysis of the experimental signals that have been carried out suggests that SAMPM and SACMPM can be used for the analysis of signals coming from complex structures. The error rate obtained here with $100$ terms is not extremely low ($\simeq50$~\% for SAMPM and $\simeq20$~\% for SACMPM), but the visual results presented in Figures~\ref{fig:perfcompSAMPM_XP} and~\ref{fig:perfcompSACMPM_XP} suggest that the approximation is very good, catching the main features of the signal. As expected by the fact that SACMPM incorporates the ability to model dispersion, its convergence rate is better than that of SAMPM. In terms of physical interpretability, the results provided by the SACMPM better corroborate with the expectations that one can have regarding the first $S_0$ and $A_0$ wave packet propagation in such a structure. 

%\end{comment}

\section{Damage localization using SAMPM and SACMPM features}\label{sec:Damage_detect}

In order to demonstrate the practical usefulness of the proposed signal approximation techniques for LW-based SHM purposes, this section aims at localizing damage in a structure using features obtained after SAMPM and SACMPM decomposition of LW signals that are then fed to a neural network. 

\subsection{Numerical database of LW-based SHM signals}

\begin{figure}[!ht]
    \centering
    \includegraphics[height=4cm]{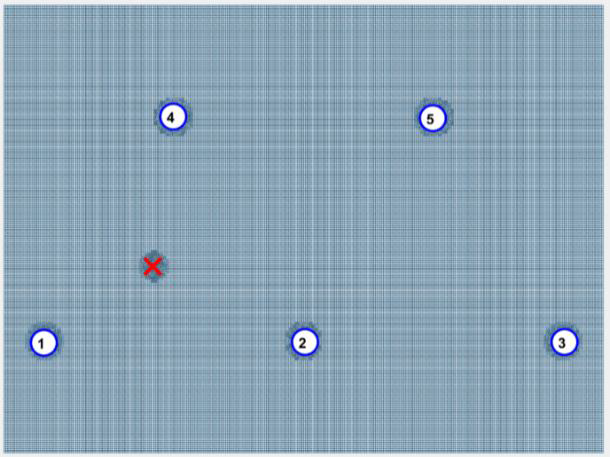}
		\includegraphics[height=3.5cm]{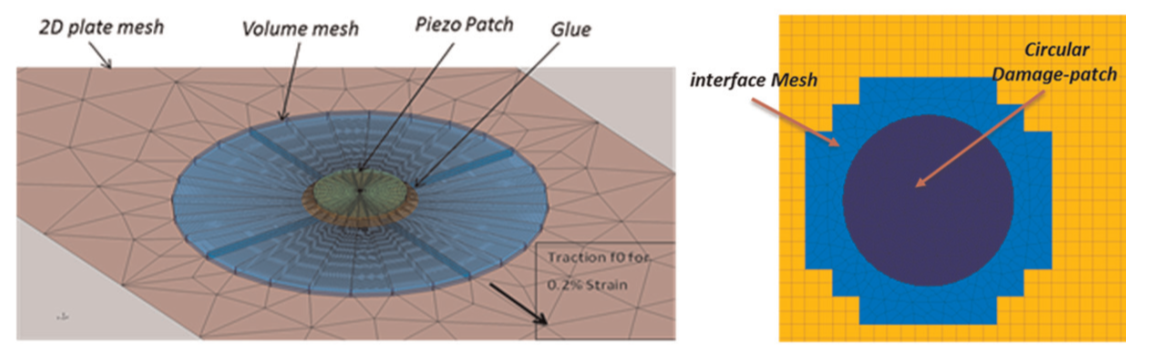}
    \caption{\emph{[Left]} Overview of the geometrical configuration for the simulated database. Mesh details around the PZT element \emph{[Center]} and the damage \emph{[Right]}.} 
    \label{fig:geometricalconfigurationsimulated}
\end{figure}

The structure under study is now a $300 \times 300 \times 2.4$~mm$^3$ plate made up of a composite material representative of the aeronautic industry. Fig.~\ref{fig:geometricalconfigurationsimulated} illustrates the disposition of PZTs as well as one damage on the simulated structure, and computational details are provided in \ref{sec:db_loc}.

\subsection{Problem statement}

A machine learning approach consisting of a feedforward neural network (NN) is considered for damage localization purposes using features extracted from LW signals using either SAMPM or SACMPM described previously. 

As input for the NN, one thus considers the constant amplitude for SAMPM or the wave packet impulse responses for SACMPM and the time delays $\lbrace \alpha_{k}(t), \tau_{k} \rbrace_{k=1}^{\PGDm}$ with $\PGDm$ the number of terms in the approximation when SAMPM or SACMPM are applied to the signal created as the difference between an \emph{undamaged} and \emph{damaged} response of the LW measured by PZTs distributed in the structure. To compare both methods, here we considered $m=6$ terms for both methods for all treated signals. As output of the NN, one considers the damage location expressed here as $\vx$. In this sense, for the  damage localization, one considers the following neural network mapping $\text{NN}(\cdot)$:
\begin{equation}
\text{NN}( \lbrace \alpha_{k}(t), \tau_{k} \rbrace_{k=1}^{\PGDm} ) \rightarrow \vx_{\text{pred}}
\end{equation}
where, $\vx_{\text{pred}}$ denotes the predicted location of damage.

The architecture considered for the neural network consists of a feed forward network, with $3$ hidden layers, where each layer has 150 neurons. The activation functions considered are hyperbolic tangent (\emph{tanh}), and the output activation function is linear. As training data, $37$ data points are considered, while $5$ are used as test data.

\subsection{Damage localization results}\label{sec:numerical_dam_loc}

\begin{figure}[!ht] %!ht
\centering
\begin{subfigure}{0.48\textwidth}
\includegraphics[width=\textwidth]{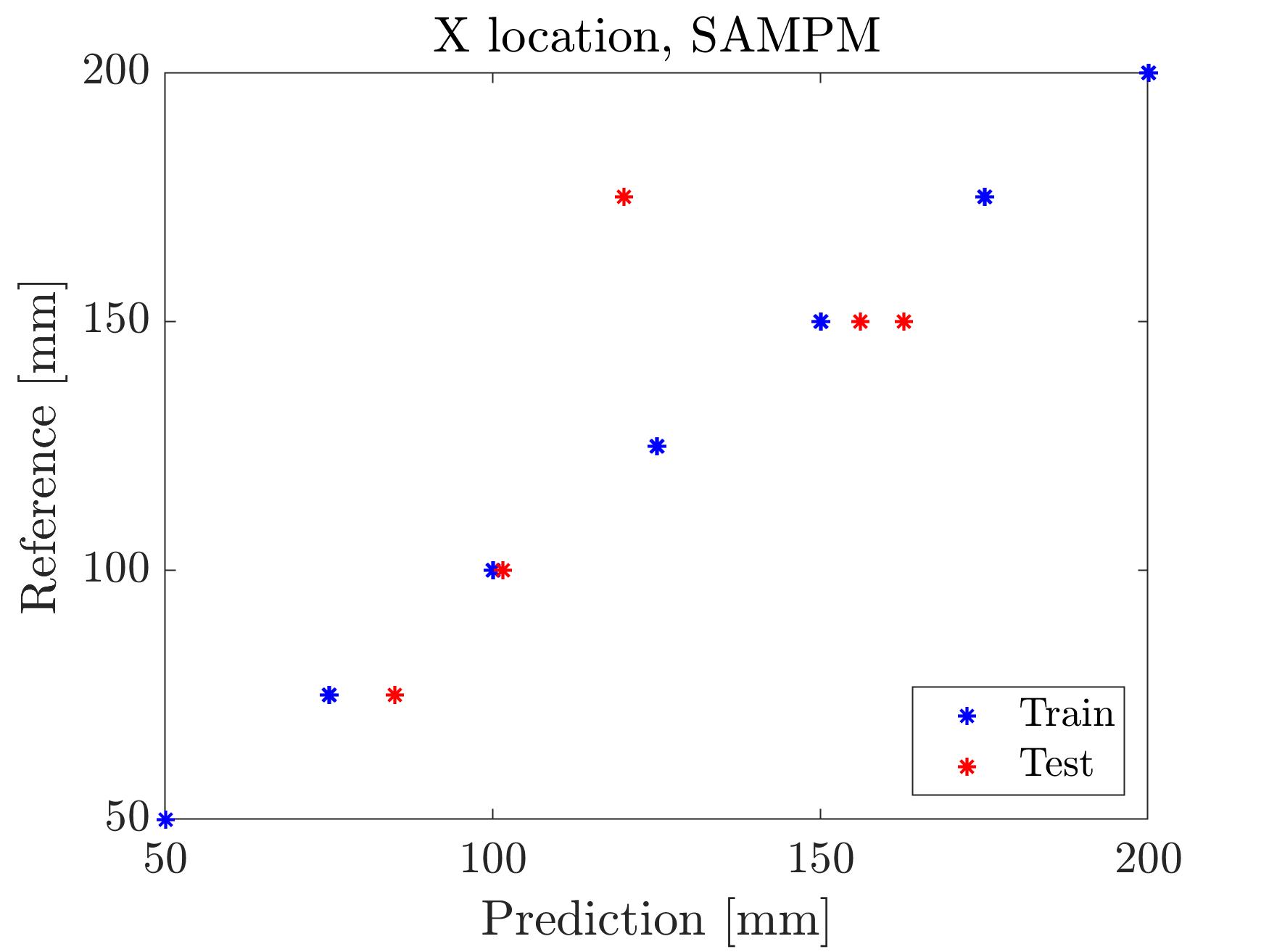}
\caption{}
\end{subfigure}
\begin{subfigure}{0.48\textwidth}
\includegraphics[width=\textwidth]{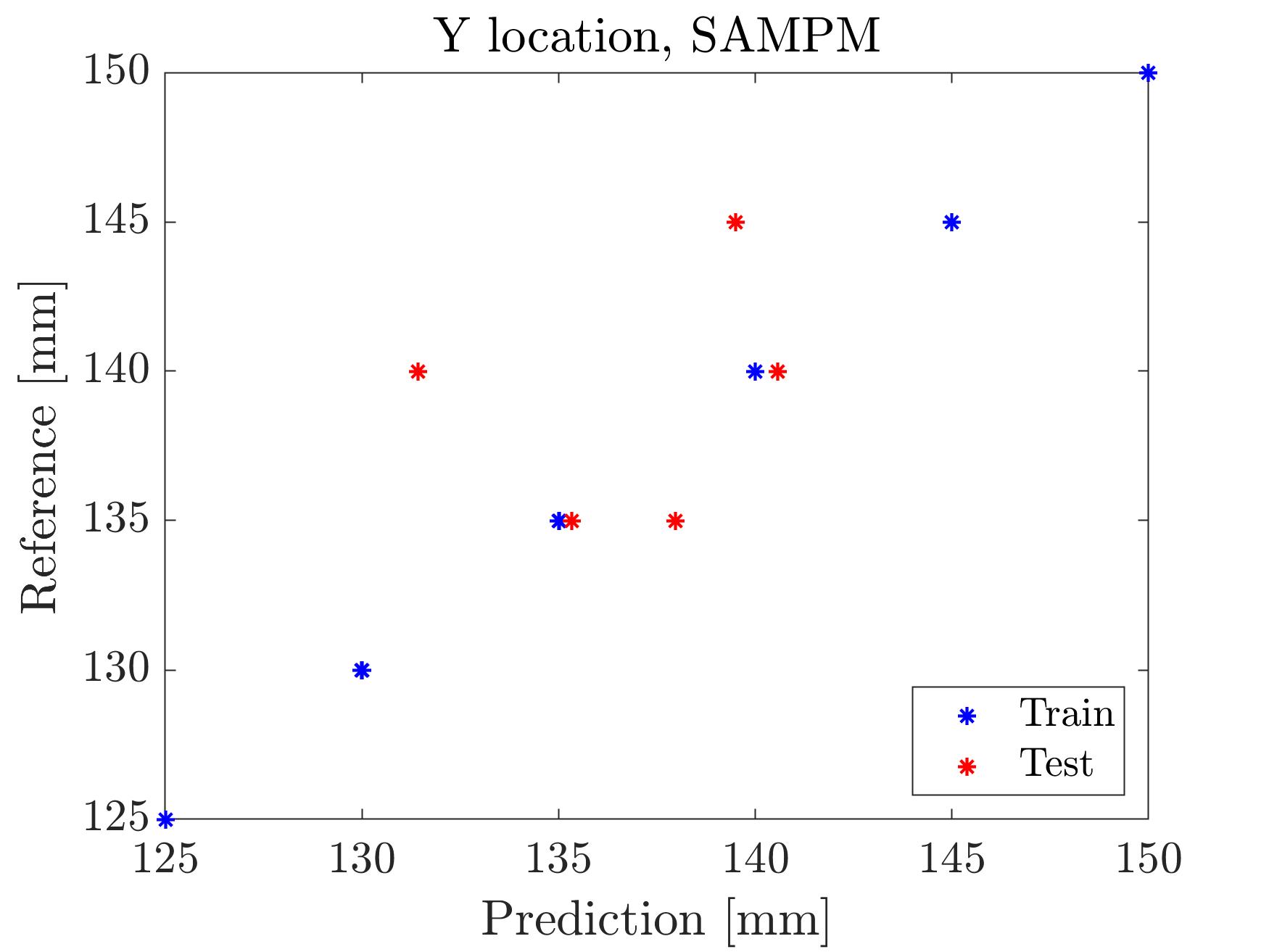}
\caption{}
\end{subfigure}
\caption{Damage location prediction for train and test data when using the SAMPM.}
\label{fig:pred_train_test_sampm}
\end{figure}

\begin{figure}[!ht] %!ht
\centering
\begin{subfigure}{0.48\textwidth}
\includegraphics[width=\textwidth]{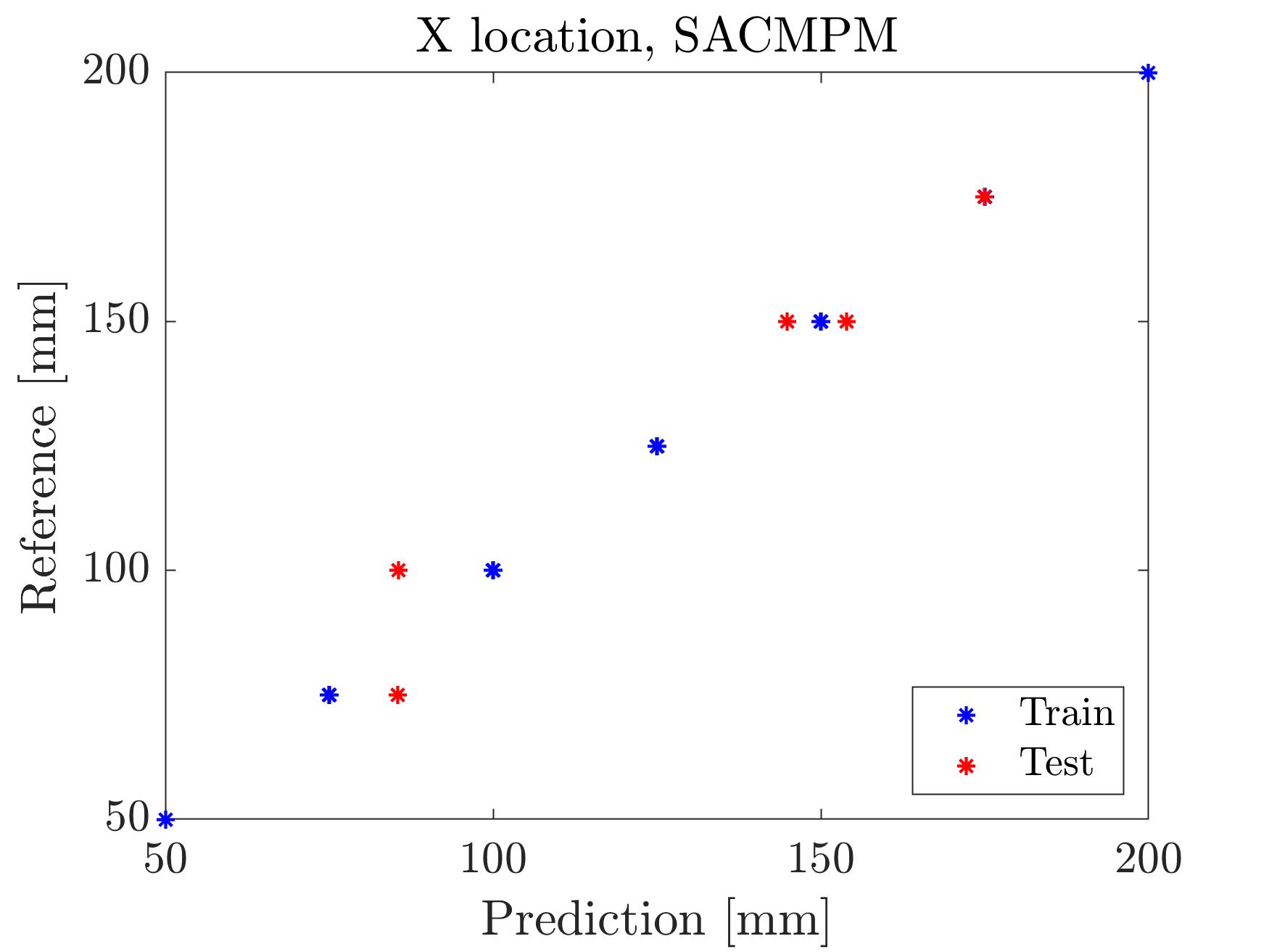}
\caption{}
\end{subfigure}
\begin{subfigure}{0.48\textwidth}
\includegraphics[width=\textwidth]{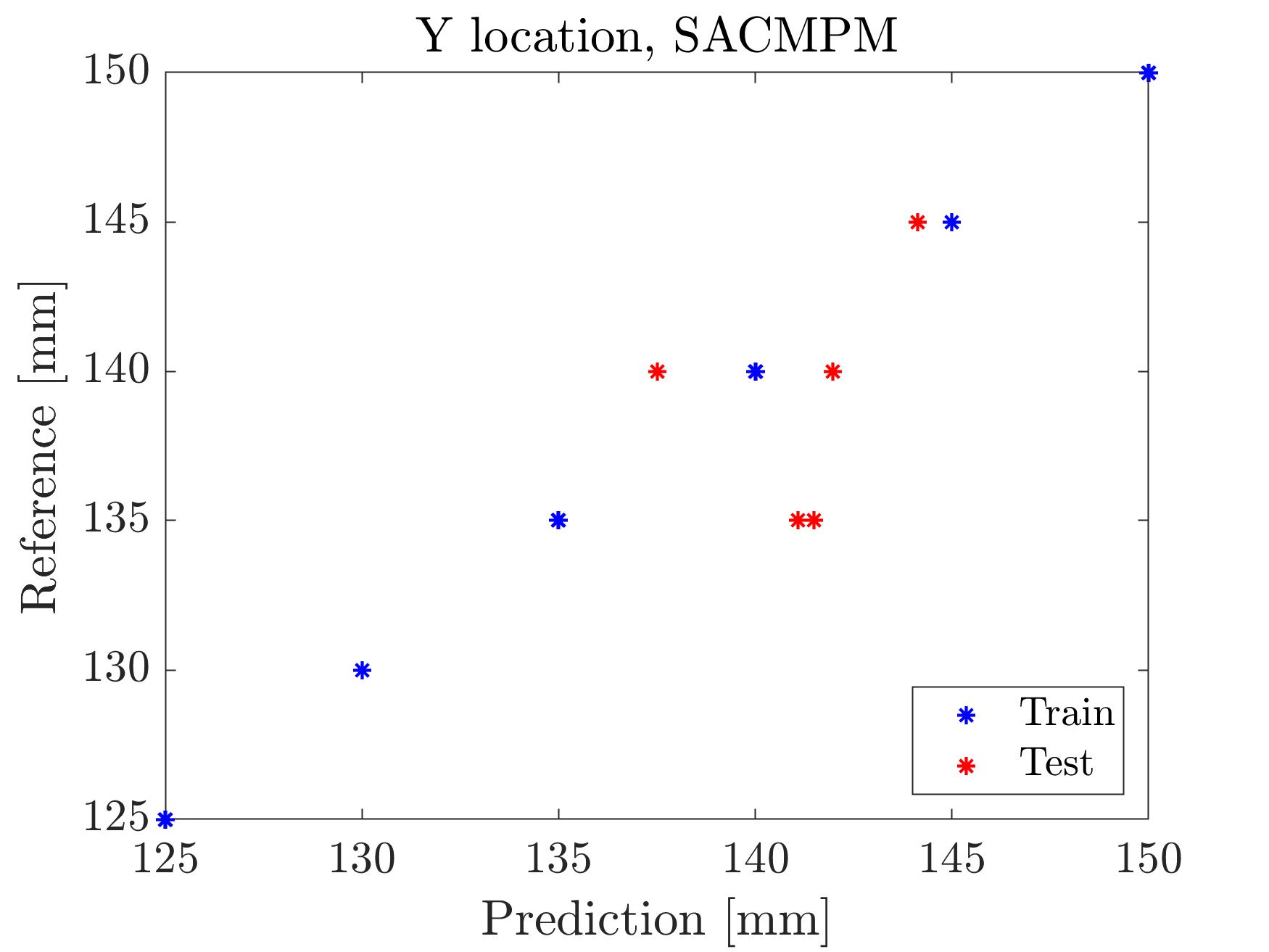}
\caption{}
\end{subfigure}
\caption{Damage location prediction for train and test data when using the SACMPM.}
\label{fig:pred_train_test_sacmpm}
\end{figure}

The prediction of damage location given by the NN for the training as well as the test data-set for the SAMPM are presented in Fig.~\ref{fig:pred_train_test_sampm}. On the other hand, the results corresponding to SACMPM are presented in Fig.~\ref{fig:pred_train_test_sacmpm}.
The prediction results of damage location in terms of relative errors of the NN by using the features of the SAMPM and SACMPM are summarized in Table~\ref{tab:NN_loc_pred}.
\begin{table}[!ht]
\centering
\begin{tabular}{| c | c | c | }
\hline
Method & Test error $x$ coordinate $[\%]$ & Test error $y$ coordinate $[\%]$ \\
\hline
SAMPM & 3.44 & 19.63 \\ 
\hline  
SACMPM & 3.18 & 13.76 \\
\hline
\end{tabular}
\caption{Error for the prediction of damage location by using the features of SAMPM and SACMPM.}
\label{tab:NN_loc_pred}
\end{table}
From the results presented in~Table \ref{tab:NN_loc_pred}, one can conclude that the SAMPM allows obtaining good features in order to produce damage identification, however, since the identification error associated with the SACMPM is lower, this means this decomposition allows obtaining richer features, that are more efficiently used by the NN for damage localization purposes.

\section{Conclusions and perspectives}\label{sec:Concl_Pers}

In the present paper, a mathematical framework is first introduced for the determination of a Single Atom Matching Pursuit Method (SAMPM) allowing to approximate Lamb waves based structural health monitoring signals without the need for a predefined dictionary. Then, this idea is extended to take into account dispersive phenomena. The proposed extension is called Single Atom Convolutional Matching Pursuit Method (SACMPM). The main idea consists of decomposing a measured signal as a delayed atom convolved with temporal functions, being richer features that represent a given Lamb wave signals to be recorded in a thin structure. The SAMPM and SACMPM methods were applied to approximate numerical signals as well as signals measured experimentally. Finally, the features extracted from both methods were also used as input to feed a neural network to predict damage location, where good predictions were obtained. 

As a perspective for future work, it is considered to provide the online calculation of atoms together with the SAMPM and SACMPM, where these atoms would allow to better approximate the reference signals and obtain in parallel an extraction of better features to improve the damage identification.

Although the signal approximation methods proposed in this paper find an original application in the context of SHM, these techniques are completely general and can be easily applied to any signal processing problem. Thus highlighting their application to various areas of engineering.

\section*{CRediT authorship contribution statement}

\textbf{Sebastian Rodriguez}: Writing – original draft, Writing – review \& editing, Visualization, Validation, Software, Methodology, Investigation, Formal analysis, Conceptualization. \textbf{Marc Rébillat}: Writing – review \& editing, Visualization, Validation, Software, 
Methodology, Conceptualization. \textbf{Shweta Paunikar}: Writing – review \& editing. \textbf{Pierre Margerit}: Writing – review \& editing, Methodology. \textbf{Eric Monteiro}: Writing – review \& editing. \textbf{Francisco Chinesta}: Writing – review \& editing, Supervision.
\textbf{Nazih Mechbal}: Writing – review \& editing, Supervision.

\section*{Declaration of competing interest}

The authors declare that they have no known competing financial interests or personal relationships that could have appeared to influence the work reported in this paper.  

\section*{Acknowledgment}

The European MORPHO project is gratefully acknowledged for funding this research activity.

\appendix

\section{Infinite plate taken as a numerical illustrative example}
\label{sec:numerical_example}

The example infinite plate used for illustration has a thickness $h=2$~mm and is made up of an isotropic material with $E=70$~GPa, $\nu=0.3$, and $\rho=1500$~kg/m$^3$. Such a plate merely corresponds to a composite plate commonly used in aeronautic structures.

In the low frequency range of such structure, only two wave modes exist, namely the $A_0$ and $S_0$ modes, and given an input frequency $f$, the corresponding wavenumbers $k_{S_0}(f)$ and $k_{A_0}(f)$ can be computed as \citep{su_guided_2006, su_identification_2009, mitra_guided_2016, qing_piezoelectric_2019}:

\begin{equation}
    k_{S_0}(f) = 2\pi f \sqrt{\frac{\rho}{Q}}
\end{equation}
\begin{equation}
    k_{A_0}(f) = \frac{2\pi f}{\sqrt{2}}
    \sqrt{
    \left(\frac{\rho}{Q}+\frac{\rho}{G\xi}\right)
    + \sqrt{
    \left(\frac{\rho}{Q} - \frac{\rho }{G\xi}\right)^2 + 
    \frac{1}{\pi f^2} \left(\rho + \frac{\rho }{I Q} \right)
    }
    }
\end{equation}
    
with: $G = \frac{E}{2(1+\nu)}$, $Q = \frac{E}{(1-\nu^2)}$, $\xi=\pi^2/12$, and $I = h^3/12$.

\begin{figure}[!ht]
\centering
    \includegraphics[width=\textwidth]{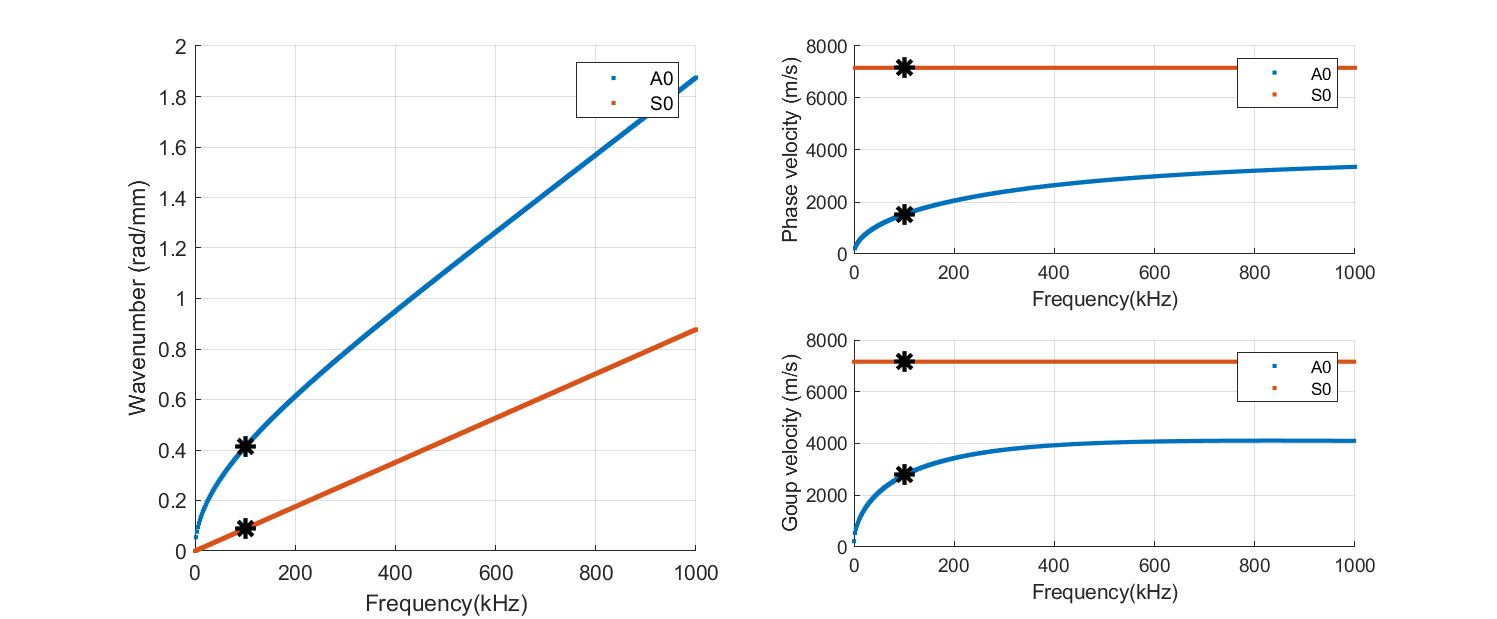}
    \caption{Dispersion curves for the simulated structure under study. [Left] Wavenumber $k$ versus frequency. [Top right] Phase velocity versus frequency. [Bottom right] Group velocity versus frequency. The black symbols denotes $A_0$ and $S_0$ waves properties at the selected input frequency of $f_0=100$~kHz.}
    \label{fig:disp}
\end{figure}

The resulting dispersion curves for the simulated structure under study are shown in Fig.~\ref{fig:disp}. On these figures, the two dispersion branches corresponding to the $A_0$ and $S_0$ modes can be seen, as well as the phase and group velocities. The phase velocity of the $S_0$ mode does not change with frequency, meaning that the $S_0$ mode does not endure dispersion, whereas the one of the $A_0$ mode is enduring dispersion. Furthermore, at the selected input frequency of $f_0=100$~kHz, it can clearly be seen that the structure is excited in a frequency region where dispersion is important for the $A_0$ mode.

Given an input signal $x(t)$ as the one shown in Fig.~\ref{fig:input} (or equivalently its Fourier transform $\hat{x}(f)$), the propagated signal for a propagation distance $d$ can be computed as:
\begin{equation}
 s_d(t) = \mathcal{F}^{-1}\left[\displaystyle \sum_{n=A_0,S_0} \int_{-\infty}^{+\infty} \hat{x}(f) exp\left[-i k_n(f) d\right] df\right]
 \label{eq:propTH}
\end{equation}
where $\mathcal{F}^{-1}$ denotes the inverse Fourier transform. Please notice that in Eq.~\eqref{eq:propTH}, it is assumed that the $S_0$ and $A_0$ modes are excited with an equal amplitude, which is not necessarily the case in practice due to the size of the PZT elements mainly. Signals have been computed using Eq.~\eqref{eq:propTH} for distances $d$ ranging from $15$~cm to $55$~cm by step of $5$~cm.

\section{Experimental study on a A380 fan cowl structure}
\label{sec:XP_details}

The fan cowl part of an Airbus A380 nacelle under study here is $1.5$ m in height for a semi circumference of $4$ m and is made of composite monolithic carbon epoxy material. It has been equipped with $30$ PZTs manufactured by NOLIAC (diameter of $25$ mm) and possesses many stiffeners delimiting various areas as shown in Fig.~\ref{fig:FC}.

The excitation signal sent to the PZT element is a 5 cycled burst with an excitation frequency of $f_0=200$~kHz and with an amplitude of $10$~V. The excitation frequency is selected to promote the mode $S_0$ over the mode $A_0$ as it propagates faster. The Lamb wave propagation speed within the material is estimated at around $5300$~m/s for the $S_0$ mode and $1800$~m/s for the $A_0$ mode. In each phase of the experimental procedure, one PZT is selected as the actuator, and the other acts as sensors. All the PZTs act sequentially as actuators. Resulting signals are then simultaneously recorded by the other piezoelectric elements and consist of $1000$ data points sampled at $1$~MHz. %Only PZT transducers from $18$ to $23$ will be considered here as they are more or less aligned and provide measurements in a given propagation direction for various distances. The PZT $23$ will emit the burst signal that will be afterward collected by the PZT elements ranging from $22$ to $18$ by descending order.

\section{Numerical database for damage localization}
\label{sec:db_loc}

To build up the numerical database, a $[0^{\circ}/45^{\circ}/ 23^{\circ}/0^{\circ}]$ composite laminate where the mechanical properties of each ply described in Tab.~\ref{tab:Mechanicalpropertiesofply} is considered. A set of 5 piezoelectric elements (Noliac NCE51), each with a diameter of 20 mm and a thickness of 0.1 mm, are surface-mounted on the composite plate. An illustration of the plate and sensor placement is shown in Fig.~\ref{fig:geometricalconfigurationsimulated}. Numerical simulations are conducted using SDTools~\citep{SDTools}. Squared elements with dimension 2 mm $\times$ 2 mm were used for the meshing. The time step for the transient simulation is chosen as 0.3 ms and leads to a sampling frequency of 3.33 MHz. The damage has a circular shape with a 20 mm diameter. The damage is represented by a local reduction in material properties of $90$~\% in the damaged area. Damage cases encompass all combination of damage position with $x=50, 75, 100, 125, 150, 175, 200$~mm and $y=125, 130, 135, 140, 145, 150$~mm. 

This FEM model was previously validated through experiments~\citep{Fendzi2016}. After the simulation, a white Gaussian noise is added to the simulation results for each path between a given actuator and a given sensor in order to simulate experimental noise. Several realizations of this noise constitute an equivalence to the experimental repetitions. A central frequency of $f_0=200$~kHz is used with SNR values equal to $150$~dB. Here, SNR stands for Signal to Noise Ratio, and the value $0$~dB refers to the maximum amount of noise pollution (the energy of the noise being equal to the energy of the signal).
\begin{table}[htb]
    \centering
    \scriptsize
    \begin{tabular}{| l | l | l | l | l | l | l |}
    \hline
    Density (g/$m^3$) & Thickness (mm) & $E_{11}$ (GPa) & $E_{22}$ & $E_{33}$ (GPa) & $G_{12}$ (GPa) & $v_{12}$ \\
    \hline
    1554 & 0.28 & 60 & 40 & 8.1 & 4.8 & 0.03 \\ \hline
    \end{tabular}
    \caption{Mechanical properties of one ply of the chosen composite material}
    \label{tab:Mechanicalpropertiesofply}
\end{table}

%%       \cite{<label>} ==> [1]

%%
%Example citation, See \cite{lamport94}.

\bibliographystyle{elsarticle-num} 
\bibliography{references}



\end{document}